\documentclass[aps,prl,twocolumn]{revtex4}

\usepackage{natbib}

\usepackage{graphicx}

\usepackage{hyperref}

\newcommand{\re}[1]{(\ref{#1})}

\renewcommand{\j}{{\protect\raisebox{1pt}[0pt][0pt]{\protect\scriptsize $j$}} }

\newcommand{\beg}{\begin{equation}}
\newcommand{\en}{\end{equation}}

\newcommand{\eps}{\epsilon}

\begin{document}

\title{Relaxation and persistent oscillations of the order parameter in the non-stationary BCS theory}

\author{Emil A. Yuzbashyan$^{1}$ }

\author{Oleksandr Tsyplyatyev$^{2}$}

\author{Boris L. Altshuler$^{3,4}$}

\affiliation{ \phantom{a}
\vspace{0.1cm}
\centerline{$^1$Center for Materials Theory,  Rutgers University,
Piscataway, New Jersey 08854, USA}
\centerline{$^2$Physics Department, Lancaster University, Lancaster LA1 4YB, UK}
\centerline{$^3$Physics Department, Columbia University,
New York, New York 10027, USA}
\centerline{$^4$NEC-Laboratories America, 4 Independence Way, Princeton, New Jersey 085540, USA}
}




\begin{abstract}
We determine the limiting dynamics of a fermionic condensate following a sudden perturbation for various
initial conditions. We demonstrate that possible initial states of the condensate fall into two classes. In the first case, the order parameter asymptotes to a {\it constant} value. The approach
to a constant is oscillatory with an inverse square root decay. This happens, e.g., when the strength of pairing
is abruptly changed while the system is in the paired ground state and more generally for any {\it nonequilibrium}  state that is in the same class as the ground state. In the second case, the order parameter
exhibits persistent oscillations with several  frequencies. This is realized for nonequilibrium states that belong to
the same class as excited stationary states. Our {\it classification} of initial states extends the concept of excitation spectrum
to nonequilibrium regime and allows one to predict the evolution without solving equations of motion.

\end{abstract}

\maketitle

The response of a fermionic condensate to fast external perturbations presents a long-standing
problem[\citealp{ander1}--\citealp{warner}]. The main difficulty is to
describe the time evolution in the non-adiabatic regime when a nonequilibrium state
of the condensate is created on a time scale shorter than the Cooper instability time $\tau_\Delta=1/\Delta_0$, where $\Delta_0$ is the equilibrium BCS gap. In this case the
evolution of the system cannot be described in terms of a quasiparticle spectrum or a
single time-dependent order parameter
$\Delta(t)$ \cite{kopnin}. One has to account for the dynamics of individual Cooper pairs, making
it a complex many-body problem.

The non-adiabatic regime can be accessed experimentally in ultra-cold Fermi gases, where the strength of pairing  between fermions can be rapidly changed\cite{gases}. Non-adiabatic measurements can be also performed
in quantum circuits utilizing superconducting qubits (nanoscale superconductors). Here nonequilibrium conditions can be generated by fast voltage pulses
on a time scale comparable to $\tau_\Delta$\cite{qubits}.

Here we consider a BCS condensate that is out of  equilibrium at $t=0$ and study its
time evolution for $t>0$. Given the state of the system at $t=0$,
we  predict the dynamics with no need for actually solving
equations of motion. We show that physically meaningful initial  states fall into two
broad categories. In the first scenario, $|\Delta(t)|$ asymptotes to a constant value $\Delta_\infty<\Delta_0$. The approach to
$\Delta_\infty$ is  oscillatory with a $1/\sqrt{t}$ decay,
\beg
\frac{|\Delta(t)|}{\Delta_\infty}=1 + a\frac{\cos(2\Delta_\infty t+\phi)}{\sqrt{\Delta_\infty t} },
\label{asymp}
\en
where the constants $a$ and $\phi$ depend on details of the initial state.
  This scenario is realized, for example, when the pairing strength
is abruptly changed,  while the system is in the paired ground state.
In the second scenario,   $|\Delta(t)|$ oscillates persistently with several incommensurate
frequencies.
The number of frequencies as well as the limiting dynamics of individual pairs can be inferred directly from the initial state.

We propose a topological  {\it classification} of initial states, which extends the concept of
excitation spectrum to the nonequilibrium regime. If a state is in the same class as the paired ground state,
Eq.~\re{asymp} applies. Other states are topologically distinct, in which case persistent
oscillations occur.

Our approach  explains  differences between   previous studies of  condensate dynamics.
 A linear analysis around the BCS ground state  yields\cite{volkov} damped oscillations with a frequency $2\Delta_0$. Eq.~\re{asymp} generalizes this result to the nonlinear case and a wide range of initial conditions.
 An oscillatory decay  following a change in
the coupling strength was observed numerically\cite{amin,simons}. We will see that this  is due to the fact that  initial states of Refs.~\cite{volkov,amin,simons} are in the same class as the BCS ground state.
Undamped periodic oscillations of $|\Delta(t)|$ have been found in Refs.~\cite{galperin,shumeiko,barankov}. They were also seen numerically for initial
states close to a normal state\cite{barankov2}.
In contrast, Ref.~\cite{warner} also starts from the normal state, but obtains
a saturation to $\Delta_\infty=\Delta_0/2$. It turns out\cite{emil3}  that this  occurs if
the initial state is a {\it paired} state with a small seed gap $\Delta_{in}\ll\Delta_0$.
 Quasiperiodic oscillations of the order parameter\cite{emil,emil1} can also be realized  (see below).

In the non-dissipative regime, one can use the BCS model to describe the dynamics of the condensate.
Here we are interested in the thermodynamic limit.
Then,  the BCS mean-field is valid as long as the order parameter is nonzero.
 There are several  equivalent ways
to derive  mean-field evolution equations.
Using Anderson's pseudospin representation\cite{ander1}, one can describe the mean-field evolution  by a {\it classical} spin Hamiltonian\cite{emil}
\beg
H_{BCS}=\sum_j 2\eps_j s_j^z-g\sum_{j,k} s_j^+s_k^-,
\label{class}
\en
where $\eps_j$ are the single-particle energies and $s_j^\pm=s_j^x\pm is_j^y$.
Dynamical variables ${\bf s}_j$ are vectors of fixed length, $|{\bf s}_j|=1/2$.
 The BCS order parameter  is $\Delta(t)=\Delta_x-i\Delta_y=g\sum_j s_j^-$.
Mean-field equations of motion   $\dot {\bf s}_j=\{H_{BCS}, {\bf s}_j\}$  follow from Hamiltonian \re{class} and the angular momentum Poisson brackets,
$\left\{ s_j^x, s_k^y\right\}=- \delta_{jk} s_j^z$ etc.,
\beg
\dot {\bf s}_j={\bf b}_j \times {\bf s}_j \phantom{m} {\bf b}_j=\left(-2\Delta_x, -2 \Delta_y, 2\eps_j\right).
\label{spins2}
\en
The general non-stationary BCS problem   can now be formulated as follows: given an initial distribution of  spins, determine its evolution.

Solution of Eq.~\re{spins2} for each of the dynamical variables ${\bf s}_j(t)$ and $\Delta(t)$ is a superposition of harmonic oscillations with certain frequencies.
In the thermodynamic limit the frequency spectrum  in general consists of  continuum and discrete parts.
Consider, e.g., the absolute value of the order parameter $|\Delta(t)|$. Suppose the discrete part  contains $k$ incommensurate frequencies. Let  their contribution to  $|\Delta(t)|$ be $F_k(t)$,
\beg
|\Delta(t)|= F_k(t)+\int_{-D}^D d\omega A(\omega)\cos[\omega t+\phi(\omega)],
\label{fewthermo}
\en
where $D$ is the ultraviolet cutoff (Debye frequency in conventional BCS). The integral in Eq.~\re{fewthermo} vanishes as $t\to\infty$ for any well behaved $A(\omega)$. Hence,
$|\Delta(t)|\to F_k(t)$ for $t\to\infty$,
i.e. $|\Delta(t)|$  oscillates with $k$  frequencies for $k\ge 1$ and
$|\Delta(t)|\to\Delta_\infty$ if $k=0$.

Given the initial state, the structure of the frequency spectrum can be determined using integrability of  BCS dynamics\cite{emil,emil1}. Frequencies of an integrable
system depend only on its integrals of motion\cite{arnold}, which can be evaluated
at $t=0$. Frequencies can thus be determined without  solving  equations of motion.

Here we use the method of Ref.~\cite{emil1}. It is convenient to introduce the following vector function (Lax vector) of an auxiliary (spectral)
parameter $u$:
\beg
{\bf L}(u)=- \frac{\hat {\bf z}}{g}+\sum_j \frac{{\bf s}_j}{u-\eps_j},
\label{lax}
\en
where $\hat {\bf z}$ is a unit vector along the $z$-axis.
The square of the Lax vector is conserved by the evolution.
  The frequency spectrum is  related to branch cuts of $w(u)=\sqrt{{\bf L}^2(u)}$. Note that the numerator of ${\bf L}^2(u)$
is a polynomial  of degree $2n$, where $n$ is the total number of levels $\eps_j$. Since ${\bf L}^2(u)\ge 0$, all $2n$ roots come in complex conjugate pairs. For finite $n$, all roots are typically distinct leading to $n$ isolated cuts connecting pairs of conjugate roots. This situation is described by the general solution\cite{emil} for finite $n$. One can also have a situation when
$2(n-m)$ roots are real and therefore double degenerate. This leaves $m$  branch cuts corresponding to $m$ pairs of complex conjugate roots. The dynamics can now be described
in terms of only $m<n$ effective spins governed by Hamiltonian \re{class} with  $m$ spins and $m$ new effective energy levels. These {\it m-spin} solutions contain only $m$ incommensurate frequencies.
One frequency corresponds to a uniform rotation of all spins around the $z$-axis. Thus, $|\Delta(t)|$   contains $m-1$ frequencies~\cite{emil1}.

In the thermodynamic limit some roots of  ${\bf L}^2(u)$ merge into continuous lines and give rise to the continuum part of the spectrum, while isolated pairs of roots correspond to the discrete part. {\it Thus, the number
of discrete frequencies in $|\Delta(t)|$ is  the number of isolated pairs of roots less one, $k=m-1$}. At large times
$|\Delta(t)|$ exhibits persistent oscillations with $k$  frequencies\cite{ph} and is described  by an $m$-spin solution\cite{mspin}.  The number $k$ is a topological property of the initial state.   It is the number of
handles on the Riemann surface of the function $\sqrt{{\bf L}^2(u)}$.

Discrete part of the frequency spectrum turns out to be related to discontinuities of the spin distribution ${\bf s}(\eps)$
as a function of $\eps$.
To see this,  consider  first stationary states.
There are two types of such states.  The BCS ground state and excited states with   a constant $\Delta\ne 0$ are  obtained by aligning each spin ${\bf s}_j$ self-consistently along its effective magnetic field ${\bf b}_j$. These states can be termed ${\it anomalous}$ stationary states. Choosing the $x$-axis so that $\Delta$ is real, we obtain
\beg
2s_j^z=-\frac{e_j\eps_j}{ \sqrt{\eps_\j^2+\Delta^2} } \quad 2s_j^x=-\frac{e_j\Delta}{ \sqrt{\eps_\j^2+\Delta^2} },
\label{ex}
\en
where $e_j=1$ if the spin is antiparallel to the field and $e_j=-1$ otherwise.
The self-consistency condition $\Delta=g\sum_j s_j^x$ yields the BCS gap equation. The state with all $e_j=1$   is the BCS ground state. The state  ${e_k=-1}$ and $e_{j\ne k}=1$   is a state with a single excited
pair\cite{ander1} of energy
$2\sqrt{\eps_k^2+\Delta_0^2}$. Using Eqs.~(\ref{lax}, \ref{ex}) and the gap equation, we derive
\beg
{\bf L}_{s}(u)=-(\Delta\hat{\bf x}+u\hat{\bf z})L_s(u),
\label{laxBCSex}
\en
\beg
L_s(u)=\sum_j \frac{e_j}{2(u-\eps_j)\sqrt{\eps_\j^2+\Delta^2} }
\label{l0bcsex}
\en
We see that ${\bf L}_{s}^2(u)=(u^2+\Delta^2) L_s^2(u)$ has a pair of isolated roots at $u=\pm i\Delta$.
All other roots are determined by the equation $L_s(u)=0$ and are double degenerate.

\begin{figure*}[p]
\includegraphics[scale=0.5,angle=0]{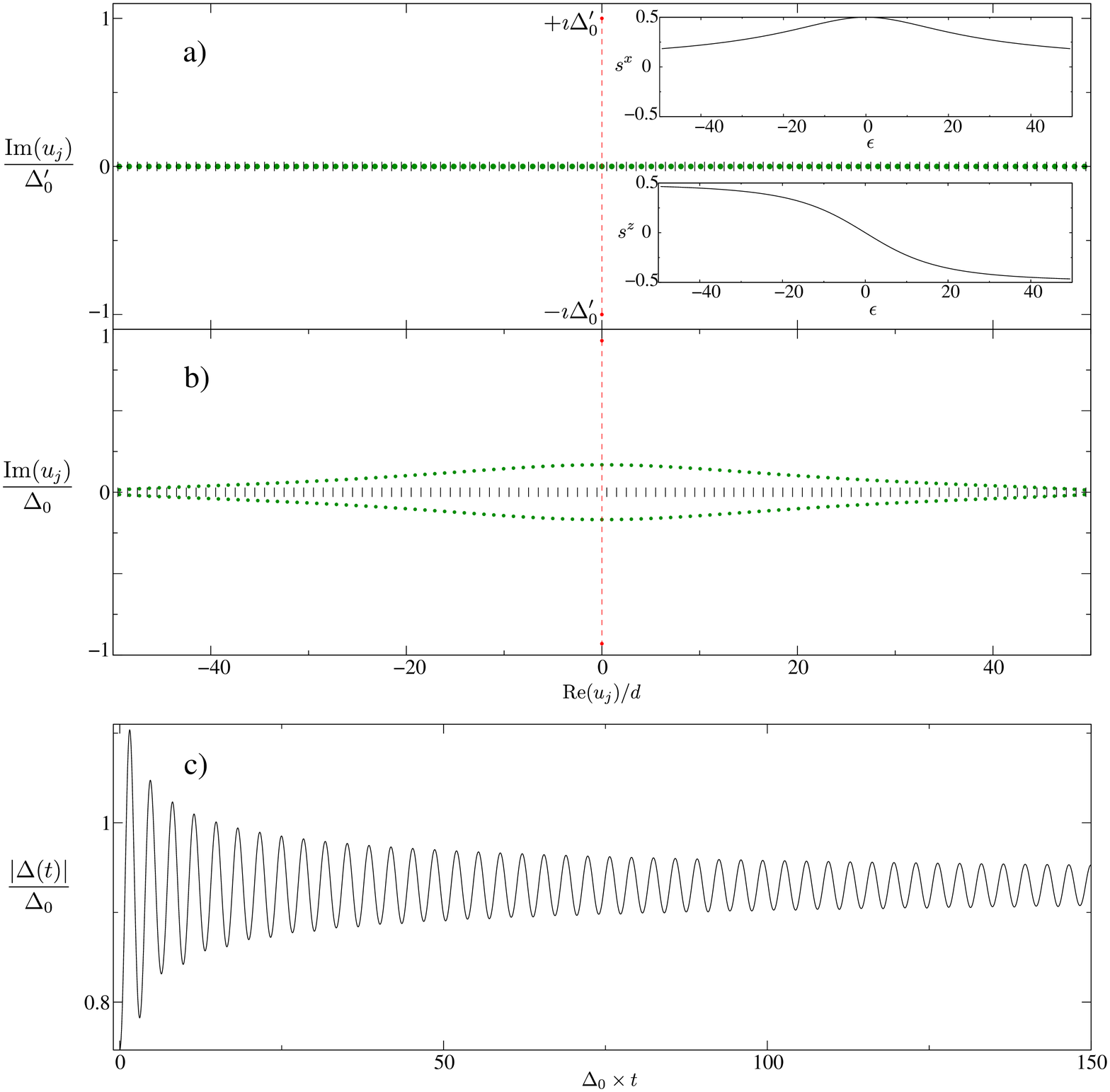}
\caption{ {\bf a)} Roots  $u_j$ of ${\bf L}^2(u)$ for the BCS ground state of $n=100$ spins with equilibrium gap $\Delta'_0$.  Axes:  $\mbox{Im} (u_j)/\Delta'_0$ and $\mbox{Re} (u_j)/d$, where
$d$ is the level spacing. Insets:   spin distribution
$s^z(\eps)$ and $s^x(\eps)$ ($s^y(\eps)=0$).  The absence of discontinuities in the spin distribution
($2p=0$) implies a single pair of isolated  roots at $\pm i\Delta'_0$ (red circles connected by a dashed line). Remaining $2n-2=198$ roots (solid green circles) are real, doubly degenerate, and located between
consecutive  energy levels $\eps_j$ (ticks on the real axis).\\
\\
{\bf b)} At $t=0$ the coupling constant is abruptly increased so that the corresponding ground state gap is $\Delta_0=2.4\Delta'_0$, while the spin
configuration remains the same as in part a).  The line of double real roots deforms into two  complex conjugate lines. The Riemann surface of $\sqrt{{\bf L}^2(u)}$ has $m=2p+1=1$
isolated cuts (red circles connected by a dashed line). This
nonequilibrium state has the same number of isolated roots as the BCS ground state and, therefore,
is in the same topological class. The frequency
spectrum of $|\Delta(t)|$ is thus continuous: $k=m-1=2p=0$, i.e. no discrete frequencies in Eq.~\re{fewthermo}.\\
\\
 {\bf c)} Time evolution of $|\Delta(t)|$ after the change  $\Delta'_0\to \Delta_0$.   In the absence of discrete frequencies, the order parameter
 $|\Delta(t)|$ asymptotes to a constant value $\Delta_\infty$ (see
 Eq.~\re{asymp}).} \label{1cut}
\end{figure*}

\begin{figure*}[p]
\includegraphics[scale=0.5]{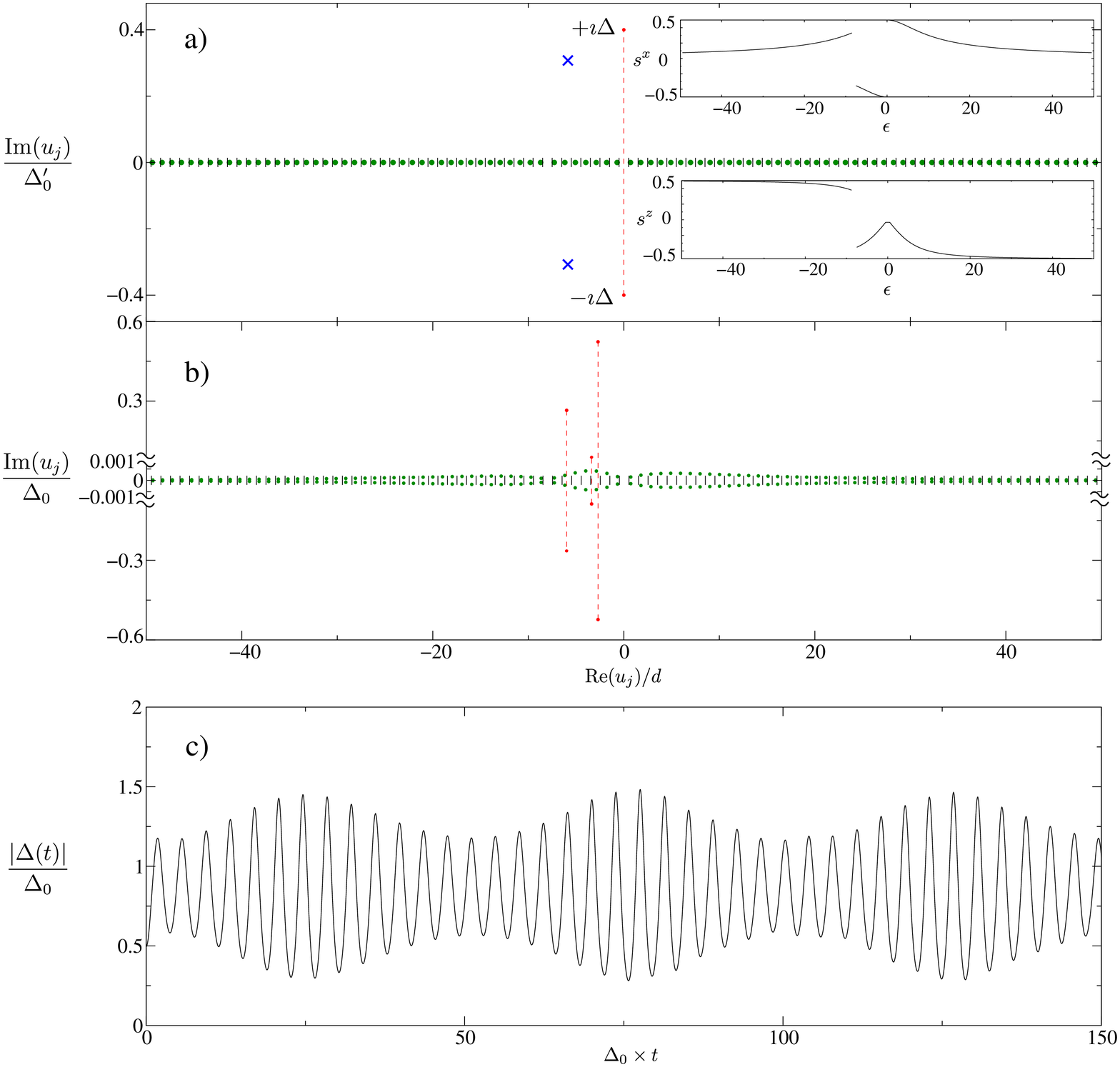}
\caption{ {\bf a)} Roots $u_j$ of ${\bf L}^2(u)$ of $n=100$ spins in an excited stationary state. This {\it anomalous} state (see the text) has been
obtained from the  ground state with the gap $\Delta'_0$ by flipping spins in energy interval $(-0.37\Delta'_0, 0)$. It has a  gap $\Delta=0.4\Delta_0'$. Axes: $\mbox{Im} (u_j)/\Delta'_0$ and $\mbox{Re} (u_j)/d$, where
$d $ is the  level spacing.    Insets: spin distribution
$s^z(\eps)$ and $s^x(\eps)$ ($s^y(\eps)=0$). Spin flips result in $2p=2$  discontinuities in the spin distribution.
Therefore, there is a $p=1$ pair of isolated doubly degenerate roots (blue crosses) in  addition to a pair of roots  $u_j=\pm i\Delta$ (red circles connected by a dashed line).  The remaining $2n-6=194$ roots (solid green circles) are real, doubly degenerate, and located between
consecutive  energy levels $\eps_j$ (ticks on the real axis).\\
\\
{\bf b)} At $t=0$ the coupling constant is abruptly increased so that the corresponding ground state gap is $\Delta_0=1.55\Delta'_0$, while the spin
configuration remains the same as in  part a).
The $p=1$ pair of isolated double roots splits into $2p=2$ pairs of isolated roots. Together with a pair of roots
coming from $u_j=\pm i\Delta$, there are $m=2p+1=3$ pairs of isolated roots (red circles connected by  dashed lines). The line of double real roots deforms into two complex conjugate lines. The Riemann surface of $\sqrt{{\bf L}^2(u)}$ has $m=2p+1=3$
isolated cuts. This nonequilibrium
 state has the same number of isolated roots as the excited state in part a) and therefore
is in the same topological class. The frequency
spectrum of $|\Delta(t)|$ has $k=m-1=2p=2$  discrete frequencies.  \\
\\
{\bf c)} Time evolution of $|\Delta(t)|$ after the change $\Delta'_0\to \Delta_0$.  Since there are $k=m-1=2p=2$ discrete frequencies
in Eq.~\re{fewthermo}, the order parameter
 $|\Delta(t)|$ displays oscillations with two  basic frequencies. Its functional form is described by an ($m=3$)--spin solution (see the text
below Eq.~\re{lax}).} \label{3cuts}
\end{figure*}

First, consider the ground state,
$e_j=1$. Since $L_s(u)\to\pm\infty$ as $u\to\eps_j\pm0$ for each $j$,  all roots of $L_s(u)$ are real and located between consecutive $\eps_j$.
In the thermodynamic limit they merge into a  line from $-D$ to $D$ as shown in Fig.~\ref{1cut}a.  Note that
the existence of a double real root between $\eps_j$ and $\eps_{j+1}$ relies on $e_j=e_{j+1}$. Further,  Eq.~\re{ex} implies that when
$e_j=e_{j+1}$, the components of spins ${\bf s}(\eps)$ are continuous at $\eps=\eps_j$.
Now let $e_j=-e_{j+1}$. In this case the real root between $\eps_j$ and $\eps_{j+1}$ can disappear.
Thus, discontinuities (jumps) in the  spin distribution  generate isolated complex
roots (see
Fig.~\ref{3cuts}a for an example).  Since  spins
far from the Fermi level are not flipped, the total number of jumps is even. In general, for $2p$ jumps
$L_s(u)$ can have up to $p$ pairs of isolated roots.

One can study equations of motion \re{spins2} linearized around anomalous stationary states. Setting
$\delta{\bf s}_j={\bf A}_je^{i\omega_k t}$, we solve for the normal modes. The eigenvalues
turn out to be $\omega_j=2\sqrt{u_\j^2+\Delta^2}$, where $u_j$ are the roots of ${\bf L}^2_{s}(u)$. For the ground state $x_j=\eps_j$ up to finite size corrections (cf. Ref.~\cite{ander1}).

Next,  consider few examples of initial states far from equilibrium. Let the system be in an anomalous stationary
state for $t<0$. Suppose at $t=0$ the coupling
changes abruptly from $g'$ to $g$.  Using Eq.~\re{lax}, one can show that the change in $g$ results
in a smooth deformation of the root distribution. Lines of roots  deform into lines.  On the other hand, doubly degenerate roots become non-degenerate.
A state that had $p$ pairs of degenerate isolated roots in addition to a pair of roots $\pm i\Delta$ now has $m=2p+1$ pairs of non-degenerate roots,
i.e. $2p+1$ cuts of $\sqrt{{\bf L}^2(u)}$.
As shown above, in this case $|\Delta(t)|$ exhibits persistent oscillations with
$k=m-1=2p$ frequencies. This behavior is illustrated in Fig.~\ref{3cuts}c.

  Let the initial state be the ground state with coupling
$g'$. Then, the line of double real roots splits into two complex conjugate lines (Fig.~\ref{1cut}b).
There is only one pair of isolated roots as in the ground state. Therefore, $k=0$ and $|\Delta(t)|$ {\it asymptotes
to a constant} $\Delta_\infty$ at $t\gg\tau_\Delta$, as illustrated in Fig.~\ref{1cut}c.
According to Eq.~\re{spins2}, at large times  spin ${\bf s}_j$ rotates in a constant
magnetic field ${\bf b}_j=(2\Delta_\infty, 0, 2\eps_j)$ with a frequency $\omega(\eps_j)=2\sqrt{\eps_\j^2+\Delta_\infty^2}$. Using this, one derives Eq.~\re{asymp}. The $1/\sqrt{t}$ decay law is set by the square root singularity in the spectral density\cite{emil3}. Note that, even though $|\Delta(t)|$ asymptotes to a constant, the final state of the system is
 non-stationary.
In the final state each spin precesses with its own frequency\cite{warner}.

\begin{figure*}[p]
\includegraphics[scale=0.5,angle=0]{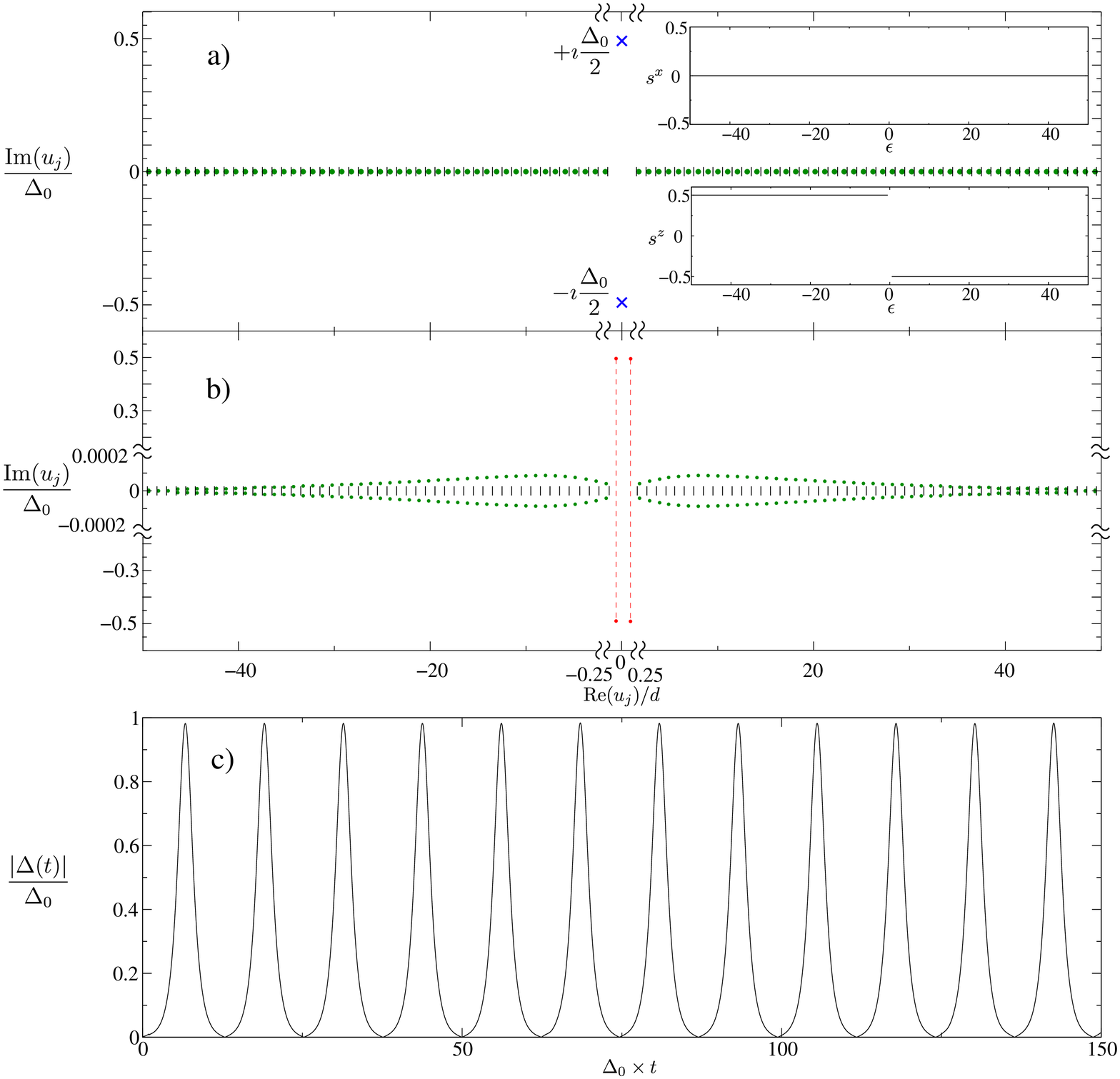}
\caption{{\bf a)} Roots $u_j$ of ${\bf L}^2(u)$ for the Fermi  state of $n=100$ spins with ground state gap $\Delta_0$.  Axes: $\mbox{Im} (u_j)/\Delta_0$ and $\mbox{Re} (u_j)/d$, where 
$d$ is the level spacing.
Insets:  spin distribution
$s^z(\eps)=-\mbox{sgn}\,\eps/2$ ($s^x(\eps)=s^y(\eps)=0$). There is $2p+1=1$ discontinuity (a jump in $s^z(\eps)$ at the Fermi level) in the spin distribution resulting in $p+1=1$ pair of imaginary doubly degenerate roots $u_j=\pm i \Delta_0/2$ (blue crosses). Remaining $2n-4=196$ roots (solid green circles) are real, doubly degenerate and located between
consecutive energy levels $\eps_j$ (ticks on the real axis).     Since all spins are 
along the $z$-axis, this state is stationary (see Eq.~\re{spins2}). The normal frequencies  are $\omega_j=2u_j$. The  mode
 corresponding to $u_j=\pm i \Delta_0/2$ is unstable, it grows as $e^{\Delta_0 t}$.  \\
\\
{\bf b)}  Initially the system is in the Fermi  state. A small ``external magnetic field'' $-2\Delta_{QF}\hat{\bf x}=-5.4\times 10^{-2}\Delta_0\hat{\bf x}$   is added to Eq.~\re{spins2}  for a time $t^*=1/\Delta_0$. The $p+1=1$ pair of double roots $u_j=\pm i \Delta_0/2$ splits into $m=2p+2=2$ pairs of isolated roots (red circles connected by  dashed lines); the line of real roots
splits  into two  complex conjugate lines. This  nonequilibrium state has the same number of isolated roots as the Fermi  state and, therefore, belongs to   the same topological class. The Riemann surface of $\sqrt{{\bf L}^2(u)}$ has $m=2p+2=2$
 isolated cuts. The frequency
spectrum of $|\Delta(t)|$ has $k=m-1=2p+1=1$ discrete frequency  in Eq.~\re{fewthermo}.
\\
\\
{\bf c)} Time evolution of $|\Delta(t)|$ for the initial state described in part b). Since there is one discrete frequency,
$|\Delta(t)|$ exhibits periodic oscillations. Its functional form is described by an ($m=2$)--spin solution (see the text
below Eq.~\re{lax}).} \label{2cuts}
\end{figure*}

 There is another type of stationary states -- {\it normal} states.
 In these states  each spin is aligned along
the $z$-axis, $s_j^z=z_j/2=\pm 1/2$.  The Fermi state is $z_j=-\mbox{sgn}\,\eps_j$ (levels below the Fermi energy are occupied, above -- empty). States with other $z_j$ correspond to particle-hole excitations of the Fermi gas. For
example, a state $z_k=\mbox{sgn}\,\eps_k<0$
has a pair of fermions removed from the level $\eps_k$.
The
Lax vector for normal states is ${\bf L}_n(u)=L_n(u)\hat{\bf z}$, where
\beg
L_n(u)=-\frac{1}{g}+\sum_j\frac{z_j}{2(u-\eps_j)}
\label{l0}
\en
All roots of ${\bf L}^2(u)=L_n^2(u)$ are thus doubly degenerate. Note the absence of a branch cut
of $\sqrt{{\bf L}^2(u)}$ connecting the points $u=\pm i\Delta$. Further analysis is similar to
that for anomalous states. The Fermi  state has a single jump in the $s_z(\eps)$ at the Fermi level.
This results in a pair of complex conjugate isolated  roots (Fig.~\ref{2cuts}a), which
 can be determined from the equation $L_n(u)=0$. In the thermodynamic limit, we obtain $u=\pm i\Delta_0/2$. The rest of the roots are real and form a line from $-D$ to $D$. For a general normal state with $2p+1$ jumps\cite{norm} in $s_z(\eps)$, one
can have up to $p+1$ complex conjugate pairs of roots. Each root is doubly degenerate. Linearizing Eqs.~\re{spins2} around a normal state, we  obtain normal  frequencies $\omega_j=2u_j$, where $L_n(u_j)=0$.
  In particular, the Fermi  state has a single unstable mode that corresponds
to $u_j=\pm i\Delta_0/2$. This mode grows  as $e^{\Delta_0 t}$ indicating the pairing instability of the Fermi state\cite{elihu,schmid}. Remaining frequencies are real and correspond to the precession of spins at their
natural frequencies, $\omega_k=2\eps_k$ up to finite size corrections.

Let the system be in or close to a normal state at $t=0$. First, we consider the Fermi  state. Since, within mean-field, normal states are unstable equilibria, a small perturbation is needed to start off the
dynamics.  One can start e.g. from a non-stationary spin distribution close to the Fermi  state\cite{barankov2}.
A typical deviation  splits all double degenerate roots as illustrated in Fig.~\ref{2cuts}b.  Real roots split into two complex
conjugate lines  close to the real axis.   Their contribution to $|\Delta(t)|$  is small in the deviation from the Fermi state and decaying.
Degenerate complex
roots at  $u=\pm i\Delta_0/2$ split into $m=2$  isolated
cuts close to each other. Since $k=m-1=1$ in Eq.~\re{fewthermo}, $|\Delta(t)|$ will exhibit undamped periodic oscillations, as shown in Fig~\ref{2cuts}c. Its functional
form is described by a 2-spin solution\cite{mspin}. The  roots close to the real axis indicate that the initial spin distribution does not match the 2-spin solution exactly (see the discussion of $m$-spin solutions above).

The dynamics in normal states can also be triggered  by quantum fluctuations. In Ref.~\cite{warner} this is modelled by adding
a small {\it external} field $-2\Delta_{QF}\hat {\bf x}$ to Eq.~\re{spins2}. This violates conservation of ${\bf L}^2(u)$. We have $d{\bf L}^2(u)/dt=2\Delta_{QF} L_y/g$. Being applied for a short time $t^*$, the external field  drives the system out of the normal state. Treating the evolution of ${\bf L}^2(u)$  perturbatively,
we find that the new positions of the roots are determined by the equation $L_n(u)=\pm 2i \Delta_{QF}t^*/g$.  Degenerate complex roots split  along the real axis into two cuts by  $2\Delta^2_{QF} t^*d/g$, where
$d=\langle \eps_{j+1}-\eps_j\rangle$ is the level spacing.  The resulting root configuration shown in Fig.~\ref{2cuts}b contains two cuts. Thus, $k=1$ and
 $|\Delta(t)|$ exhibits periodic oscillations, see Fig.~\ref{2cuts}c.

We see that for initial conditions close to the Fermi  state  a periodic solution is
``dynamically selected''\cite{barankov2}. Other initial conditions ``select''  damped oscillations or
multi-frequency undamped oscillations. In  more conventional terms, this
corresponds to a basic fact that the evolution of an integrable system depends on initial conditions.
All diferent behaviors are captured by the general solution\cite{emil}. Here we systematically classified possible initial states
and specialized the general solution to each type of initial conditions, i.e. we developed ``selection rules''
for the BCS dynamics.
Which behavior is realized in a particular experimental setup  depends on  the initial state of the condensate. In this respect, the periodic solution is somewhat special: if we start from the
ground state with a small $\Delta_{in}\ll \Delta_0$, the order parameter $|\Delta(t)|$ asymptotes to a constant value
$\Delta_\infty$, see Eq.~\re{asymp}. In the thermodynamic limit,  the damping disappears only when $\Delta_{in}=0$\cite{merge}.

\begin{figure*}[p]
\includegraphics[scale=0.5,angle=0]{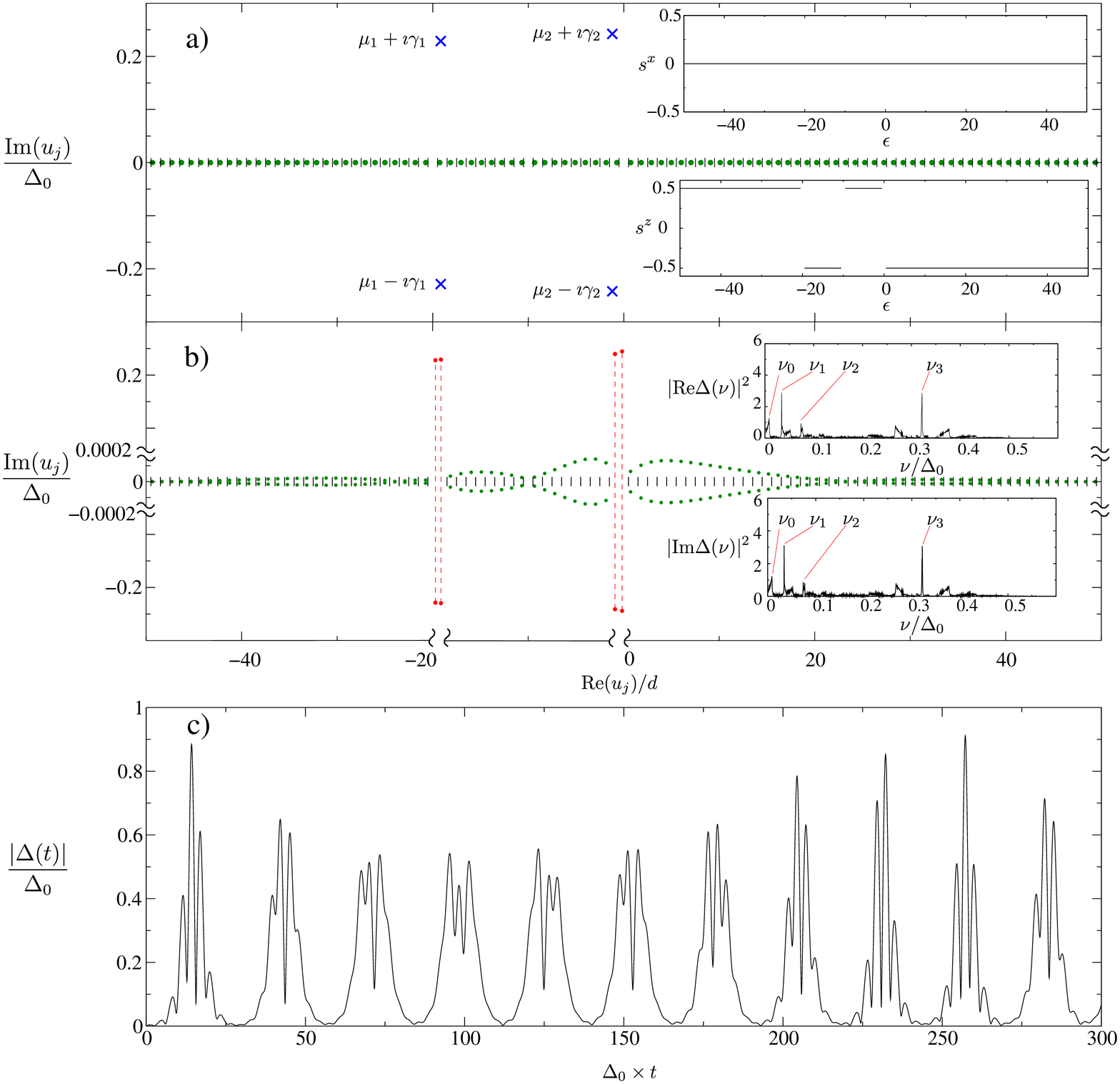}
\caption{ {\bf a)} Roots $u_j$ of ${\bf L}^2(u)$  for an excited stationary state of  $n=100$ spins with ground state gap $\Delta_0$.  Axes: $\mbox{Im} (u_j)/\Delta_0$ and $\mbox{Re} (u_j)/d$, where  
$d$ is the level spacing.
This {\it normal}  state (see the text) has been
obtained from the Fermi  state by flipping spins in energy interval $(-0.5\Delta_0, -\Delta_0)$. Insets:  spin distribution
$s^z(\eps)$ ($s^x(\eps)=s^y(\eps)=0$). There are $2p+1=3$ discontinuities (three jumps in $s^z(\eps)$) in the spin distribution resulting in $p+1=2$ pairs of isolated double roots $u_j=\mu_{1,2}\pm i \gamma_{1,2}$ (blue crosses). Remaining $2n-8=192$ roots (solid green circles) are real, doubly degenerate and located between
consecutive energy levels $\eps_j$ (ticks on the real axis).     Since all spins are 
along the $z$-axis, this state is stationary (see Eq.~\re{spins2}). The normal frequencies  are $\omega_j=2u_j$. There are two unstable modes
 corresponding to  $u_j=\mu_{1,2}\pm i \gamma_{1,2}$, which grow as $e^{2\gamma_1 t}$ and $e^{2\gamma_2 t}$.\\
 \\
{\bf  b)}  Initially the system is in the stationary state  described in part a). A small ``external magnetic field'' $-2\Delta_{QF}\hat{\bf x}=-5.4\times 10^{-2}\Delta_0\hat{\bf x}$   is added to Eq.~\re{spins2}  for a time $t^*=1/\Delta_0$.  
The $p+1=2$ pairs of degenerate roots  $u_j=\mu_{1,2}\pm i \gamma_{1,2}$ split into $m=2p+2=4$ pairs of isolated roots (red circles connected by a dashed line), while the line of real roots
splits  into two  complex conjugate lines. This nonequilibrium state has the same number of isolated roots as the stationary state in part a) and, therefore, belongs to
 the same topological class. The Riemann surface of $\sqrt{{\bf L}^2(u)}$ has $m=2p+2=4$ isolated cuts.    Insets: Fourier spectra of $\mbox{Re}\Delta(t)$
and $\mbox{Im}\Delta(t)$ display $m=2p+2=4$  basic frequencies. One frequency ($\nu_0$) corresponds to a uniform rotation of all spins around the $z$-axis and cancels out in $|\Delta(t)|$. Thus, the frequency
spectrum of $|\Delta(t)|$ has $k=m-1=2p+1=3$ discrete frequencies.
\\
\\
{\bf c)} Time evolution of $|\Delta(t)|$ for the initial state described in part b). Since  $k=m-1=2p+1=3$ 
in Eq.~\re{fewthermo},
$|\Delta(t)|$ exhibits quasiperiodic oscillations with three basic frequencies (see  insets in Fig.~\ref{4cuts}b). Its functional form is described by an ($m=4$)--spin solution (see the text
below Eq.~\re{lax}).}
\label{4cuts}
\end{figure*}

Excited normal states  have several jumps in the  spin distribution and can therefore display oscillations with more than
one frequency. Consider, e.g., a  state where  pairs in energy interval from $-\eps_a$ to $-\eps_b$ have been removed.
  The initial spin distribution is $s_z(\eps)=-\mbox{sgn}[\eps(\eps+\eps_a)(\eps+\eps_b)]$.   The $2p+1=3$ jumps\cite{norm} result in $p+1=2$ pairs of isolated double degenerate complex  roots. As a small perturbation   splits these roots into $m=2p+2=4$ cuts, $|\Delta(t)|$ oscillates with $k=m-1=2p+1=3$ incommensurate frequencies, see Fig.~\ref{4cuts}.

In conclusion, we have shown how the non-stationary BCS dynamics can be predicted from the initial state of the condensate. We classified   initial states by their integrals of motion --  roots of ${\bf L}^2(u)$. For states with a root diagram as in the paired ground state, the order parameter $|\Delta(t)|$ displays damped oscillations described by Eq.~\re{asymp}. Other states
are of the same type as excited stationary states of the BCS Hamiltonian. In these cases $|\Delta(t)|$  oscillates persistently
with few incommensurate frequencies. The number of frequencies is related to the number of jumps in the  pseudospin
distribution of the corresponding stationary state. For the situation most relevant to the experiments on cold fermions -- an abrupt change of the coupling -- we predict damped oscillations with $1/\sqrt{t}$ decay.

We thank M. Dzero, V. I. Falko, S. P. Novikov, and G. L. Warner for  discussions. This work was supported by NSF 
DMR--0210575 and DARPA under the QuIST program.

\begin{thebibliography}{99}
\bibitem{mspin} An $m$-spin solution can be expressed in terms of a Riemann theta function of genus $m-1$\cite{emil}. In particular, 2-spin solutions are
genus one -- ordinary elliptic functions.
\bibitem{ph}
If the initial state is particle-hole symmetric
$k$ is equal to the integer part of $m/2$.
In this case the effective $m$-spin problem  is also particle-hole symmetric, which reduces the number of independent degrees of freedom and therefore the number of incommensurate frequencies.
\bibitem{norm} Far from the Fermi level
$s_z(\eps)=-\mbox{sgn}\eps$. Therefore, the total number of jumps in $s_z(\eps)$ for normal states is always odd.
For anomalous stationary states the number of discontinuities in the spin distribution
is  even by the same argument.
\bibitem{merge} The two solutions merge for $\Delta_{in}$  comparable to the mean level spacing $d$.
Here we consider the thermodynamic limit $d/\Delta_{in}\to 0$.
\end{thebibliography}
\end{document}